\documentclass[11pt]{article}
\usepackage[latin1]{inputenc}
\usepackage{amsmath}
\usepackage{amsfonts}
\usepackage{amssymb}
\begin{document}
\begin{center}
{\Large \bf On Conformal d'Alembert-Like Equations\footnote{To celebrate one century of 
Special Theory of Relativity.}}
\end{center}

\bigskip
\bigskip
\begin{center}
{\bf E. Capelas de Oliveira}\\
Departamento de Matem\'atica Aplicada\\
Imecc $-$ C.P. 6065 $-$ Unicamp\\
13083-859 Campinas (SP) Brazil\\
E-mail: capelas@ime.unicamp.br\\
and\\
{\bf R. da Rocha}\\
Instituto de F\'{\i}sica``Gleb Wataghin"\\
IFGW $-$ C.P. 6165 $-$ Unicamp\\
13083-970 Campinas (SP) Brazil\\
E-mail: roldao@ifi.unicamp.br
\end{center}

\bigskip

\begin{abstract}
Using conformal coordinates associated with projective conformal 
relativity we obtain a conformal Klein-Gordon partial differential 
equation. As a particular case we present and discuss a conformal `radial'
d'Alembert-like equation. As a by-product we show that this `radial' equation 
can be identified with a one-dimensional Schr\"odinger-like equation in which the potential is exactly the second P\"oschl-Teller potential.
\end{abstract}

\section*{Introduction}
After one of the most important of Einstein's\footnote{The year 1905 is known as the \emph{ Annus Mirabilis}, because in it were published three of the most important papers of the century; \emph{Complete Papers on Light Quanta}, \emph{Brownian Motion}, and \emph{Special Theory of Relativity}.} papers \cite{eins} concerning Special Relativity was published, several alternative theories were proposed. Among them, some different interpretations and particular generalizations have been presented. In this paper we are interested in one such theory, namely the theory of hyperspherical universes, proposed by Arcidiacono\cite{arci1} several years ago and, more specifically, in the so-called conformal case.

When we write Maxwell equations in six dimensions, with six projective coordinates (we have, in these coordinates, a Pythagorean metric) a natural problem arises, namely, to provide a physical version of the formalism, i.e. to ascribe a physical meaning to the coordinates. For  this theory, there are two possible different physical interpretations: a bitemporal interpretation and a biprojective interpretation. In the first case (bitemporal) we introduce a new universal constant $c'$ and the coordinate $x_5 = i c't'$ where $t'$ is interpreted as a second time; we thus obtain in cosmic scale the so-called multitemporal relativity, proposed by Kalitzen\cite{kali}. The set of Maxwell equations obtained in this theory generalizes the equations of the unitary theory of electromagnetism and gravitation, as proposed by Corben\cite{corb}.

On the other hand (our second, biprojective case) we can interpret the extra coordinate, 
$x_5$, as a second projective coordinate. We then obtain the so-called conformal 
projective relativity, proposed by Arcidiacono\cite{arci2}, which extends in cosmic scale the theory proposed by Ingraham\cite{ingr}, but with a different physical interpretation. In this theory we have another universal constant, $r_0$, which can be taken as $r/r_0 = N$, where 
$r$ is the radius of the hypersphere and $N$ is the cosmological number appearing in the  Eddington-Dirac theory\cite{eddi}.

Here we consider only the second alternative, i.e., the biprojective interpretation. With this 
aim we introduce a projective space $P_5$ tangent to the hypersphere $S^4$. We then introduce six projective coordinates ${\overline x}_a$, with $a=0,1,\ldots,5$ and normalized 
as 
$$
{\overline x}^2 + {\overline x}^2_0 - {\overline x}^2_ 5 = r^2,
$$
where ${\overline x}^2 = x_i x^i$, $i=1,\ldots,4$ and $r$ is the radius of the hypersphere. These coordinates allow us to construct the conformal projective relativity, using a six-dimensional tensor formalism.

This paper is organized as follows: in Section 1 we present a review of the so-called theory of hyperspherical universes, proposed by Arcidiacono, considering only the six-dimensional 
case, in which conformal projective relativity appears as a particular case. The choice 
of convenient coordinates and the link between the derivatives in these two formulations (a geometric version, six-dimensional, and a physical version, the five-dimensional conformal version) are also presented. After this review, we discuss in Section 2 a Klein-Gordon partial differential equation written in conformal coordinates. In Section 3, we show that a conformal `radial' d'Alembert-like equation, can be led into a Schr\"odinger differential equation in which the associated potential is exactly a second P\"oschl-Teller potential.

\section{Hyperspherical Universes}

In 1952 Fantappi\'e proposed the so-called theory of the universes\cite{fant}.
This theory is based on group theory and on the hypothesis that the universe is endowed with unique physical laws, valid for all observers. As particular cases, Arcidiacono\cite{arci1} 
studied a limitation of that theory, i.e., he considered hyperspherical universes with 
$3,4,\ldots,n$ dimensions where motions are given by $n(n+1)/2$-parameter rotation group in spaces with $4,5,\ldots,{\mbox{(}}n+1{\mbox{)}}$ dimensions, respectively. Those models of hyperspherical $S^3, S^4,\ldots,S^n$ universes, can be interpreted 
as successive physical improvements, because any one of them (after $S^4$) contains its precedents and is contained in its successors.

After 1955 Arcidiacono studied the case $n=4$, special projective relativity, based on the 
de Sitter hyperspherical universe with a group (the so-called Fantappi\'e-de Sitter group) of ten parameters. This theory is an improvement (in a unique way) of Einstein's special relativity theory and provides a new group-theoretical version of the big-bang cosmology. As a by-product of special projective relativity one can recover several results, for example, Kinematic Relativity, proposed by Milne\cite{miln}; Stationary Cosmology, proposed by Bondi-Gold\cite{bond} and Plasma Cosmology, proposed by Alfv\`en\cite{alfv}.  

Moreover, if we consider a universe $S^4$ as globally hyperspherical but endowed with a locally variable curvature, we obtain the so-called general projective relativity which was proposed and studied by Arcidiacono after 1964. This theory allows us to recover several results as particular cases, for example, the unitary theories proposed by Weyl\cite{weyl}, 
Straneo\cite{stra}, Kaluza-Klein\cite{kalu,klei}, Veblen\cite{vebl} and Jordan-Thiry\cite{jord}
and some generalizations of the gravitational field, as those proposed by Brans-Dicke\cite{bran}, Rosen\cite{rose} and Sciama\cite{scia}.

In this paper we are interested only in the case $n=5$, i.e., conformal projective relativity based on the hyperspherical universe $S^4$ and its associated rotation group, with fifteen parameters, which contains the accelerated motions. We remember that, whereas for $n=4$ we have a unitary theory (a magnetohydrodynamic field), for $n=5$ we have another unitary theory, i.e.,
the magnetohydrodynamics and Newton's gravitation. We also present the relations between Cartesian, projective and conformal coordinates and the link involving derivatives in the six- and five-dimensional formulations.

\subsection{Conformal Coordinates}
We use the notation $x_i$, ($i=1,2,3,4$) and $x_5$ for conformal coordinates and 
${\overline x}_a$, ($a=0,1,2,3,4,5$) for projective coordinates. The relations between 
these coordinates are
$$
x_i = r_0 \frac{\overline{x}_i}{\overline{x}_0 + \overline{x}_5}
\qquad {\mbox{and}} \qquad x_5 = r_0 \frac{r}{\overline{x}_0 + \overline{x}_5},
$$
which satisfy the condition
$$
x_5^2 - x^2 = r_0^2 \frac{\overline{x}_0 - \overline{x}_5}
{\overline{x}_0 + \overline{x}_5},
$$
where $x^2 = x_i x^i$, and $r_0$ and $r$ are constants. After these considerations, the transformations of the so-called conformal projective group are obtained using the quadratic form in projective coordinates
$$
{\overline x}^2 + {\overline x}^2_0 - {\overline x}^2_ 5 = r^2,
$$
decomposing the elements of the six-dimensional rotation group (with fifteen parameters) in fifteen simple rotations (${\overline x}_a, {\overline x}_b$)\cite{eco1}.
 
\subsection{Connection Between Derivatives}
Our main objective is to write down a differential equation, more precisely a Klein-Gordon-like equation, associated with conformal coordinates. We first obtain the relation 
between the six projective derivatives ${\overline \partial}_a \equiv \partial / \partial {\overline x}_a$ and the five-dimensional derivatives $\partial_i = \partial / \partial x_i$ 
and $\partial_5 = \partial / \partial x_5$. We can then write the differential 
equations in the projective formalism, with six dimensions, in physical, i.e., conformal coordinates, with five dimensions.\footnote{As we already know, in five dimensions 
we must impose a condition on space in order to account for the fact that we are aware of only four dimensions. We have the same situation here, i.e., we must impose an additional condition.}

Taking $\phi = \phi (x_i , x_5)$, a scalar field, and using the chain rule we can write
$$
\begin{array}{ccl}
\partial_i \phi &=& \left[ ({\overline\partial}_i {\overline x}_k){\overline \partial}_k + ({\overline \partial}_i {\overline x}_5){\overline \partial}_5 + ({\overline \partial}_i {\overline x}_0) {\overline \partial}_0 \right] {\overline \phi}\\
\partial_5 \phi &=& \left[({\overline \partial}_5 {\overline x}_k) {\overline \partial}_k +
({\overline \partial}_5 {\overline x}_5){\overline \partial}_5 + ({\overline \partial}_5 {\overline x}_0) {\overline \partial}_0 \right] {\overline \phi}
\end{array}
$$
with ${\overline \phi} = {\overline \phi} ({\overline x}_i , {\overline x}_5 , {\overline x}_0)$ and $ i,k=1,2,3,4.$

From now on we take $r=1=r_0$. We consider ${\overline \phi}({\overline x}_a)$ a 
homogeneous function with degree $N$ in all six projective coordinates ${\overline x}_a$. 
Using Euler's theorem associated with homogeneous function, we get
$$
\left( \overline{x}_i \overline{\partial}_i + \overline{x}_5 \overline{\partial}_5 +
\overline{x}_0 \overline{\partial}_0 \right) \phi = N \phi
$$
where $\overline{\partial}_a = \partial / \partial \overline{x}_a$ and $N$ is the degree of homogeneity of the function. 

Then, the link between the derivatives can be written as follows\cite{arci3}
$$
\begin{array}{ccl}
\overline{\partial_0} \,\overline{\phi} &=& N\displaystyle \frac{A^{+}}{x_5} \phi +
B^{-} \partial_5 \phi - x_5 x_i \partial_i \phi\\
&&\\
\overline{\partial_5} \,\overline{\phi} &=& -N\displaystyle \frac{A^{-}}{x_5} \phi
- B^{+} \partial_5 \phi - x_5 x_i \partial_i \phi\\
&&\\
\overline{\partial_i} \,\overline{\phi} &=& N\displaystyle \frac{x_i}{x_5} \phi +
x_i \partial_5 \phi +  x_5 \partial_i \phi
\end{array}
$$
where we have introduced a convenient notation
$$
2A^{\pm} = 1 \mp x^2 \pm x_5^2 \qquad {\mbox{and}} \qquad  2B^{\pm} = 1\pm x^2 \pm x_5^2.
$$
We observe that for ${\overline x}_5 = 0$ and considering ${\overline \partial}_5 {\overline \phi} = 0$ we obtain
$$
\begin{array}{ccl}
\overline{\partial}_i \overline{\phi} & = & A \partial_i \phi +
\displaystyle \frac{N}{A} x_i \phi\\
&&\\
\overline{\partial}_0 \overline{\phi} & = & - A x_i \partial_i
\phi + \displaystyle \frac{N}{A} \phi
\end{array}
$$
where $A^2 = 1 + x^2$. These expressions are the same expressions obtained in special projective relativity\cite{eco2,eco3} and provide the link between the five projective derivatives and the four derivatives in Cartesian coordinates, i.e., the relation between five-dimensional (de Sitter) universe and four-dimensional (Minkowski) universe.

\section{Conformal Klein-Gordon Equation}
In this section we use the  previous results to calculate the so-called generalized Klein-Gordon differential equation 
$$
\frac{\partial^2}{\partial \overline{x}_a^2} \Phi + {\sf m }^2
\Phi = 0
$$
where ${\sf m}^2$ is a constant and $a=0,1,\ldots ,5.$ Introducing projective coordinates 
(in this case we have a Pythagorean metric) we obtain\footnote{Hereafter we consider 
${\sf m} = m_0 c / \hbar$ where $m_0$, $c$ and $\hbar$ have the usual meanings.} 
$$
\frac{\partial^2 U}{\partial \overline{x}_i^2} + \frac{\partial^2 U}{\partial
\overline{x}_0^2}- \frac{\partial^2 U}{\partial \overline{x}_5^2}
+ {\sf m}^2 U = 0
$$
where $i=1,2,3,4$ and $U=U(\overline{x}_i , \overline{x}_0 , \overline{x}_5)$. 

Using the relations between projective and conformal coordinates 
and the link (involving the derivatives) in the two formulations we can write
$$
\left[ x_5^2 \left( \Box - \frac{\partial^2}{\partial x_5^2}
\right) + 3x_5 \frac{\partial}{\partial x_5} + N(N+5) + {\sf
m}^2 \right] u(x_i , x_5) = 0
$$
where $N$ and ${\sf m}^2$ are constants, $\Box$ is the Dalembertian operator given by
$$
\Box = \Delta - \frac{1}{c^2} \frac{\partial^2}{\partial t^2}
$$
and $\Delta$ is the Laplacian operator. This partial differential equation is the 
so-called Klein-Gordon differential equation written in conformal coordinates or 
a conformal Klein-Gordon equation.

The case ${\sf m}^2 = 0$ transforms this equation in the so-called generalized d'Alembert differential equation. Another way to obtain this differential equation is to consider 
the conformal metric in cartesian coordinates, which furnishes the so-called Beltrami metric\cite{arci1} where the d'Alembert equation appears naturally. This equation 
can also be obtained by means of the second order Casimir invariant operator\footnote{Invariant operators associated with dynamic groups furnish mass formulas, energy spectra and, in general, characterize specific properties of physical systems.} associated with the conformal group.

To solve the conformal Klein-Gordon equation, we first introduce the spherical coordinates 
$(r, \theta , \phi)$ and get
$$
\frac{\partial^2 u}{\partial r^2} + \frac{2}{r} \frac{\partial
u}{\partial r} + \frac{1}{r^2} {\cal L}u  - \frac{1}{c^2} \frac{\partial^2
u}{\partial t^2} - \frac{\partial^2 u}{\partial x_5^2} +
\frac{3}{x_5} \frac{\partial u}{\partial x_5} +
\frac{\Lambda}{x_5^2} u = 0,
$$
where we introduced $x_4 = i ct$ and defined the operator\footnote{Here $r$ is a coordinate and should not be confused with the radius of the hypersphere. Besides, it is always possible to define a Wick-rotation\cite{saku} of the time coordinate, i.e., $ct$ $\mapsto$ $ict$.} 
$$
{\cal L} \equiv  \frac{\partial^2
}{\partial \theta^2} + \cot \theta \frac{\partial }{\partial
\theta} + \frac{1}{\sin^2 \theta} \frac{\partial^2
}{\partial \phi^2}
$$
involving only the angular part. In this partial differential equation we have $u=u(r,\theta , \phi , t, x_5)$ with $\Lambda = N(N+5) + {\sf m}^2$.

Using the method of separation of variables we can eliminate the temporal and angular parts, writing 
$$
u=u(r,\theta , \phi , t, x_5) = A\, {\mbox{e}}^{inct} Y_{\ell m} (\theta , \phi) \, f(r, x_5),
$$
where $A$ is an arbitrary constant, $n > 0$, $\ell = 0,1,\ldots$ and $m = 0, \pm 1, \ldots$ with $-\ell \leq m \leq \ell$ and $Y_{\ell m} (\theta , \phi)$ are the spherical harmonics, we get 
the following partial differential equation
$$
\frac{\partial^2 f}{\partial r^2} + \frac{2}{r} \frac{\partial
f}{\partial r} - \frac{\partial^2 f}{\partial x_5^2} +
\frac{3}{x_5} \frac{\partial f}{\partial x_5} +
\frac{\Lambda}{x_5^2} f + \left[ n^2 - \frac{\ell (\ell +1)}{r^2}
\right] f = 0
$$
with $f=f(r,x_5)$. If we impose a regular solution at the origin ($r \to 0$), the solution 
of this partial differential equation can be obtained in terms of a product of two Bessel 
functions\cite{eco1}.

\section{A d'Alembert-Like Equation}
In this section we present and discuss a partial differential equation which can be 
identified to a d'Alembert-like equation, which we call a conformal `radial' d'Alembert equation. We firstly introduce a convenient new set of coordinates, then we use separation of variables and obtain two ordinary differential equations. One of them can be identified as an ordinary differential equation whose solution is a generalization of Newton's law of gravitation; the other one is identified with an ordinary differential equation similar to a one-dimensional Schr\"odinger differential equation with a potential equal to the second P\"oschl-Teller potential.  

We introduce the following change of independent variables 
$$
\begin{array}{ccl}
r & = & \rho \cosh \xi ,\\
x_5 & = & \rho \sinh \xi ,
\end{array}
$$
with $\rho >0$ and $\xi \geq 0$, in the separated Klein-Gordon equation, obtained in the previous section, and after another separation of variables we can write a pair of ordinary differential equations, namely,  
$$
\rho^2 \frac{d^2 U}{d\rho^2} - p(p+1)U = 0,
$$
where $U=U(\rho)$ and
$$
\frac{d^2 V}{d\xi^2} + (2 \tanh \xi - 3 \coth \xi)\frac{dV}{d\xi}
+ \left[ \frac{\ell (\ell +1)}{\cosh^2 \xi} -
\frac{\Lambda}{\sinh^2 \xi} - p(p+1)\right] V =0
$$
where $V=V(\xi)$ and $p$ is a separation constant.

We first discuss the differential equation in the variable $\rho$. Its general solution is given by
$$
U(\rho) = C_1 \rho^{\,-p} + C_2 \rho^{\, p+1}
$$
where $C_1$ and $C_2$ are arbitrary constants.

If we consider the case $p=1$, introducing the notation $C_1 = gM$ with $g$ and $M$ having the usual meanings, we get
$$
U(\rho) = \frac{gM}{\rho} + C_2 \, \rho^2
$$
i.e., a gravitational potential which can be interpreted as a sum of a Kepler-like potential and a harmonic oscillator potential, giving rise to the gravitational force
$$
f(\rho) = \nabla U = - \frac{gM}{x^2 - x_5^2} + 2C_2(x^2 -
x_5^2)^{1/2},
$$
with a singularity at $x=x_5$. We note that for $C_2 = 0$ we obtain an expression analogous to  Newton's law of gravitation.

Secondly, the equation in the variable $\xi$. To solve this ordinary differential equation 
we first introduce the change of dependent variable
$$
V(\xi) = \sinh^{\frac{1}{2}} \xi  \tanh \xi \, F(\xi)
$$ 
and obtain
\begin{equation}
-\frac{d^2}{d\xi^2}F(\xi) + \left[ \frac{\mu (\mu - 1)}{\sinh^2 \xi} - \frac{\ell (\ell+1)}{\cosh^2 \xi} + \left(p+ \frac{1}{2}\right)^2 \right] F(\xi ) = 0,
\end{equation}
where the parameter $\mu$ is given by a root of the following algebraic equation $\mu (\mu - 1) = N^2 + 5N + 15/4 + {\sf m}^2$.

The differential equation above can be identified with a Schr\"odinger-like differential equation in which the associated potential is given by
$$
{\cal V}_{\mu \ell} (\xi) = \frac{\mu (\mu - 1)}{\sinh^2 \xi} - \frac{\ell (\ell+1)}{\cosh^2 \xi},
$$
which is exactly the second P\"oschl-Teller potential with energy $E$ given by $E_p = -(p+1/2)^2 < 0$. The solution of this ordinary differential equation is well known and can be expressed in terms of the hypergeometric function. An algebraic treatment can be found in \cite{baru,baru1}.

We note that the first P\"oschl-Teller potential is connected with the 
study of a Dirac particle on central backgrounds associated with an anti-de Sitter 
oscillator, i.e., the transformed radial wave functions satisfy the second-order 
Schr\"odinger differential equation whose potential is exactly the first P\"oschl-Teller
potential\cite{cota}.

Finally, a particular case of Eq.(1), i.e., the case $\mu = 0$, is related to the 
anti-de Sitter static frame as shown recently by da Rocha and Capelas de Oliveira\cite{roch}.

\section*{Concluding Remarks}
In this paper we discussed the calculation of a conformal d'Alembert-like equation.
We used the methodology of projective relativity to obtain a conformal Klein-Gordon 
differential equation and, after the separation of variables, we got another partial differential equation in only two independent variables, the so-called conformal d'Alembert differential equation. Another separation of variables led to an ordinary differential equation which generalizes Newton's law of gravitation. Finally, we showed that the remaining differential equation, a `radial' differential equation, is transformed into a one-dimensional Schr\"odinger differential equation with an associated potential that can be identified exactly with the second P\"oschl-Teller potential.  

From supersymmetric quantum mechanics with periodic potentials, it can be seen that the most general periodic potentials which can be analytically solved involve Jacobi's 
elliptic functions, which in various limits become P\"oschl-Teller potentials arising in the context of Kaluza-Klein spectrum\cite{kalu}. Kaluza-Klein modes of the graviton have been widely investigated \cite{randall,hata,bram,nam}, since the original formulation of Randall and Sundrum necessarily has a continuum of Kaluza-Klein modes without any mass gap, arising from a periodic system of 3-branes. The methods and equations developed here can shed some new light in the calculation of mass gaps from a distribution of $D$-branes\cite{nam} in the context of  five-dimensional supergravity in a forthcoming paper.

A natural continuation of this calculation is to prove that all `radial' problems associated with an equation resulting from a problem involving a light cone can be led into a Schr\"odinger-like differential equation in which the potential is exactly the P\"oschl-Teller potential\cite{ECO}.

\section*{Acknowledgment}
One of us (RR) is grateful to CAPES for financial support.
We are also grateful to Dr. Geraldo Agosti for interesting and useful discussions and Dr. J. 
Em\'{\i}lio Maiorino for several suggestions about the paper.

\end{document}